\documentclass[12pt,preprint,number]{elsarticle}

\usepackage[T1]{fontenc}
\usepackage[utf8]{inputenc}

\usepackage{amsmath,amssymb}
\usepackage{graphicx}
\usepackage{upgreek}
\usepackage{booktabs}
\usepackage{float}
\usepackage{enumitem}
\usepackage{lineno}
\usepackage{hyperref}
\usepackage{siunitx}
\usepackage{geometry}
\usepackage{sectsty}

\usepackage{setspace}
\AtBeginDocument{
	\onehalfspacing
}
\sectionfont{\large\bfseries}
\subsectionfont{\normalsize\bfseries}

\graphicspath{{./}}

\begin{document}
	
	\begin{frontmatter}

		\title{Microstructural Origins of Plastic Work Partitioning during Deformation of 310S TWIP Steel}
		
		\author[inst1]{Sandra Musia{\l}\corref{cor1}}
		\cortext[cor1]{Corresponding author: smusial@ippt.pan.pl}
		\author[inst1]{Micha{\l} Maj}
		\author[inst1]{Marcin Nowak}
		
		\address[inst1]{Institute of Fundamental Technological Research, Polish Academy of Sciences,\\ Pawińskiego 5B, 02-106 Warsaw, Poland}
	
		\begin{abstract}
			The microstructural mechanisms governing the plastic work partitioning in twinning-induced plasticity (TWIP) steels remain insufficiently understood, particularly under strain localization. This study provides a crystallographic-scale interpretation of the fraction of plastic work stored during deformation of 310S TWIP steel exhibiting complex deformation mechanisms.
			
			Electron backscatter diffraction (EBSD) was used to characterize local crystallographic orientation and microtexture during uniaxial tensile deformation using two complementary approaches: tracking the same surface region at successive strain levels and analyzing regions corresponding to known local plastic strain. Deformation was initially dominated by dislocation slip, whereas twinning activity increased significantly beyond an equivalent plastic strain of $\bar{\varepsilon}^p \approx 0.3$. Progressive deformation produced pronounced lattice rotations and a dual-fiber texture comprising dominant $\langle 111 \rangle \parallel$ rolling direction ($\mathrm{RD}$) and secondary $\langle 100 \rangle \parallel \mathrm{RD}$ components.
			
			Correlation of the EBSD observations with previously quantified evolution of the energy storage rate reveals that intensified twinning and texture redistribution in strain-localized regions are accompanied by a pronounced decrease in the fraction of plastic work stored in the material. Twin-matrix refinement and progressive lattice rotation promote non-uniform strain distribution and create favorable conditions for shear-band-mediated deformation, contributing to a reduction of the energy storage rate toward zero or negative values at advanced stages of localization.
			
		\end{abstract}
		
		
	\end{frontmatter}
	
	\newpage
	
	\section{Introduction}
	\label{intro}
	
	Transformation-induced plasticity (TRIP) and twinning-induced plasticity (TWIP) phenomena have been extensively studied in high-alloy austenitic steels because they exhibit an exceptional combination of strength, ductility, and energy absorption during plastic deformation. These properties arise from deformation mechanisms beyond conventional dislocation slip. In TRIP steels, plastic deformation promotes martensitic transformation, whereas TWIP steels accommodate strain primarily through mechanical twinning. Both mechanisms sustain strain hardening, delay necking, and improve fracture resistance, making these alloys attractive for structural and energy-absorbing applications \cite{Remy1977, Ullrich2016, Karaman2000, yang2006dependence}.
	
	The stacking fault energy (SFE) is widely recognized as the primary parameter governing deformation mechanisms in face-centred cubic (FCC) alloys. Deformation twinning typically occurs for SFE values in the range of $\SIrange{20}{50}{\milli\joule\per\meter\squared}$, whereas martensitic transformation becomes dominant below approximately $\SIrange{15}{20}{\milli\joule\per\meter\squared}$ \cite{Remy1977}. Alloys with intermediate SFE exhibit complex deformation behavior involving dislocation slip, stacking faults, twinning, and, in some cases, martensitic transformation, which strongly influences strain hardening and microstructural evolution \cite{Ullrich2016, Huang2017}. In addition to chemical composition \cite{brofman1978effect, Chen2022}, deformation mechanisms in TWIP/TRIP steels are affected by temperature \cite{Grajcar2018}, strain rate \cite{lee2014effect}, grain size, and crystallographic orientation \cite{Karaman2000, GUTIERREZURRUTIA20103552, RAHMAN2015247}. Twinning activity depends strongly on grain orientation and Schmid factor distribution, while temperature-induced changes in SFE can shift the dominant deformation mode either from phase transformation to twinning \cite{Field2024} or from twinning to slip \cite{Grajcar2018}. Recent work has also highlighted the importance of unstable fault energies (USFE and UTFE) as indicators of deformation-mode selection \cite{Wang2022}. Together, these studies demonstrate that TWIP steels exhibit strongly coupled microstructural and mechanical responses during deformation.
	
	At large plastic strains, deformation texture evolution in FCC materials is strongly influenced by mechanisms producing intense lattice rotation, including strain localization and shear-band-mediated deformation. In low- and intermediate-SFE alloys, localized shear accommodates plastic incompatibility between deformation substructures and promotes crystallographic reorientation beyond that produced by slip and twinning alone. Texture transitions involving Copper, Brass, and Goss components have been linked to shear-band-associated lattice rotation at advanced deformation levels \cite{HaileYan2014, ELDANAF20002665, Paul2007}. These mechanisms are particularly relevant in TWIP steels, where deformation twinning interacts with lattice rotation during strain localization.
	
	Alongside microstructural evolution, plastic deformation involves conversion of mechanical work into stored energy and heat. Energy storage reflects the formation and interaction of lattice defects during deformation \cite{titchener1958stored}. Since the classical work of Taylor and Quinney \cite{taylor1934latent}, it has been recognized that the fraction of stored energy evolves during deformation \cite{wolfenden1971energy, chrysochoos1989plastic, oliferuk1995effect, oliferuk2009stress}. To describe instantaneous energy conversion, the energy storage rate
	\begin{equation}
		Z=\frac{de_s}{dw_p}.
	\end{equation}
	was introduced \cite{oliferuk1985energy}. The evolution of $Z$ reflects changes in active deformation mechanisms and defect structures. Experimental investigation of energy conversion during plastic deformation has been enabled by combined Digital Image Correlation (DIC) and Infrared Thermography (IRT), which allow investigation of local thermomechanical behavior \cite{li2016local}, for example during L{\"u}ders band propagation \cite{louche2001thermal, wang2017kinematic} and reconstruction of plastic work and heat sources \cite{bodelot2009experimental, chrysochoos2009fields, knysh2015determination}. In particular, the energy balance during tensile deformation of 310S austenitic steel was quantified in \cite{musial2022field}, where the evolution of plastic work, dissipated heat, and energy storage rate was determined experimentally, accounting for both thermoelastic effect and heat exchange with the surroundings.
	
	Energy storage has also been examined from complementary microstructural perspectives. In \cite{oliferuk2007plastic} plastic instability was analyzed using thermomechanical measurements combined with transmission electron microscopy (TEM), relating stored energy to the evolution of defect structures. More generally, the energy storage in slip-dominated materials is commonly interpreted within the framework of low-energy dislocation structures (LEDSs) \cite{hansen1986low,kuhlmann1987leds,kuhlmann2001q}, whereby continued deformation promotes the rearrangement of dislocations into progressively lower-energy configurations. Such interpretations relate stored energy primarily to dislocation density and organization. In materials where deformation twinning contributes significantly to plasticity, however, crystallographic reorientation and texture evolution may also influence the partitioning of plastic work, particularly under conditions of localized deformation. This was recently demonstrated for AZ31B magnesium alloy \cite{maj2026plastic}, where thermomechanical measurements revealed fundamentally different plastic work partitioning during slip- and twinning-dominated deformation, highlighting the governing role of the active deformation mechanism in energy storage. Complementary microstructural observations confirmed that these deformation modes are associated with distinct defect evolution pathways. Whether analogous relationships exist in FCC TWIP steels remains insufficiently understood.
	
	The present work addresses this gap by correlating electron backscatter diffraction (EBSD) - based characterization of crystallographic orientation and microtexture evolution with previously determined energy-storage measurements for 310S TWIP steel. The analysis focuses on the onset and development of strain localization, where pronounced lattice rotation and refinement of the twin–matrix lamellar structure occur. Although the macroscopic energy balance during tensile deformation of 310S steel was previously quantified in \cite{musial2022field}, that study was limited to thermomechanical measurements and evaluation of energy conversion parameters. The present work constitutes an independent microstructural investigation that uses those macroscopic results as a reference framework. By linking the evolution of crystallographic texture and deformation twinning with the energy storage rate, it provides a crystallographic-scale interpretation of the conversion of plastic work into stored energy in TWIP-deforming austenitic steel.
	
	\section{Material and methods}
	\label{method}
	The investigated material was 310S austenitic stainless steel, which is widely used in high-temperature applications due to its excellent oxidation resistance and mechanical stability. It is considered a candidate structural material for supercritical water-cooled reactors (SCWRs), where long-term exposure to aggressive environments imposes stringent requirements on microstructural stability \cite{Chen2022}. The chemical composition of the material is listed in Table \ref{tab:composition}.
	\begin{table}[H]
		\centering
		\caption{Chemical composition of tested material (in wt\%).}
		\begin{tabular}{l c c c c c c c c c}
			\toprule
			& C    & S     & P     & Mn   & Si   & Cr           & Ni           & N    & Fe      \\ 
			\midrule
			310S  & 0.08 & 0.015 & 0.045 & 2.0  & 0.75 & 24.0--26.0  & 19.0--22.0  & 0.11 & balance \\ 
			\bottomrule
		\end{tabular} 
		\label{tab:composition}
	\end{table}
	Owing to its high nickel content, the austenitic phase remains stable during plastic deformation and no deformation-induced martensitic transformation is expected. The SFE, estimated using the empirical relation proposed in \cite{brofman1978effect}, lies in the range of $\SIrange{35.3}{43.4}{\milli\joule\per\meter\squared}$. Accordingly, plastic deformation is governed primarily by deformation twinning accompanied by dislocation slip, characteristic of TWIP steels.
	
\begin{figure}[H]
		\centering
		\includegraphics[width=129mm]{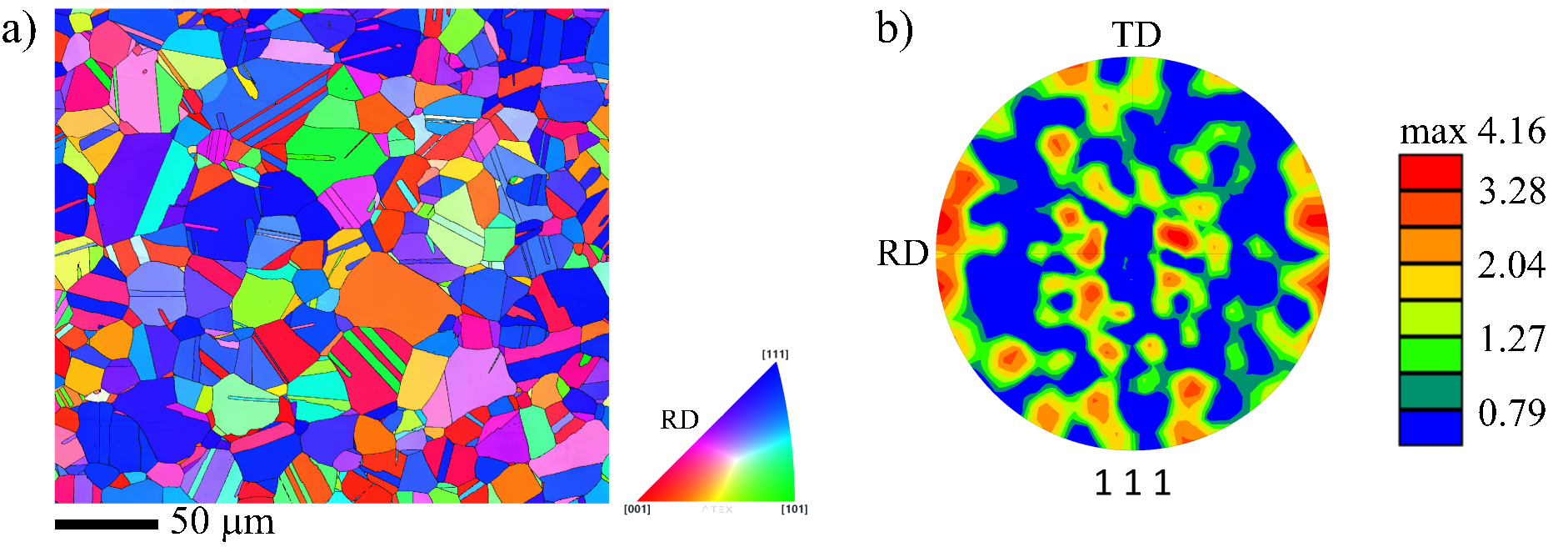}
		\caption{(a) IPF map representing grain orientations with respect to RD and (b) corresponding (111) pole figure of 310S steel in the reference state.}
		\label{fig:EBSD_initial}
	\end{figure}

	Specimens for thermomechanical testing and microstructural characterization were extracted from a cold-rolled sheet after annealing at $\SI{1373}{\kelvin}$ followed by air/water spray quenching. The initial microstructure was characterized by EBSD using a Zeiss Crossbeam 350 scanning electron microscope equipped with an EDAX Hikari Super detector. Figure \ref{fig:EBSD_initial} presents the inverse pole figure (IPF) map acquired in the RD--TD plane and transformed into the TD--ND plane, together with the corresponding (111) pole figure (RD, TD, and ND denote the rolling, transverse, and normal directions, respectively).
	The material exhibits a fully austenitic microstructure with equiaxed grains and an average grain size of approximately $\SI{21}{\micro\meter}$. A high density of annealing twins is observed, predominantly with straight and parallel morphology, while the initial crystallographic texture is weak.
	 
\begin{figure}[H]
		\centering
		\includegraphics[width=100mm]{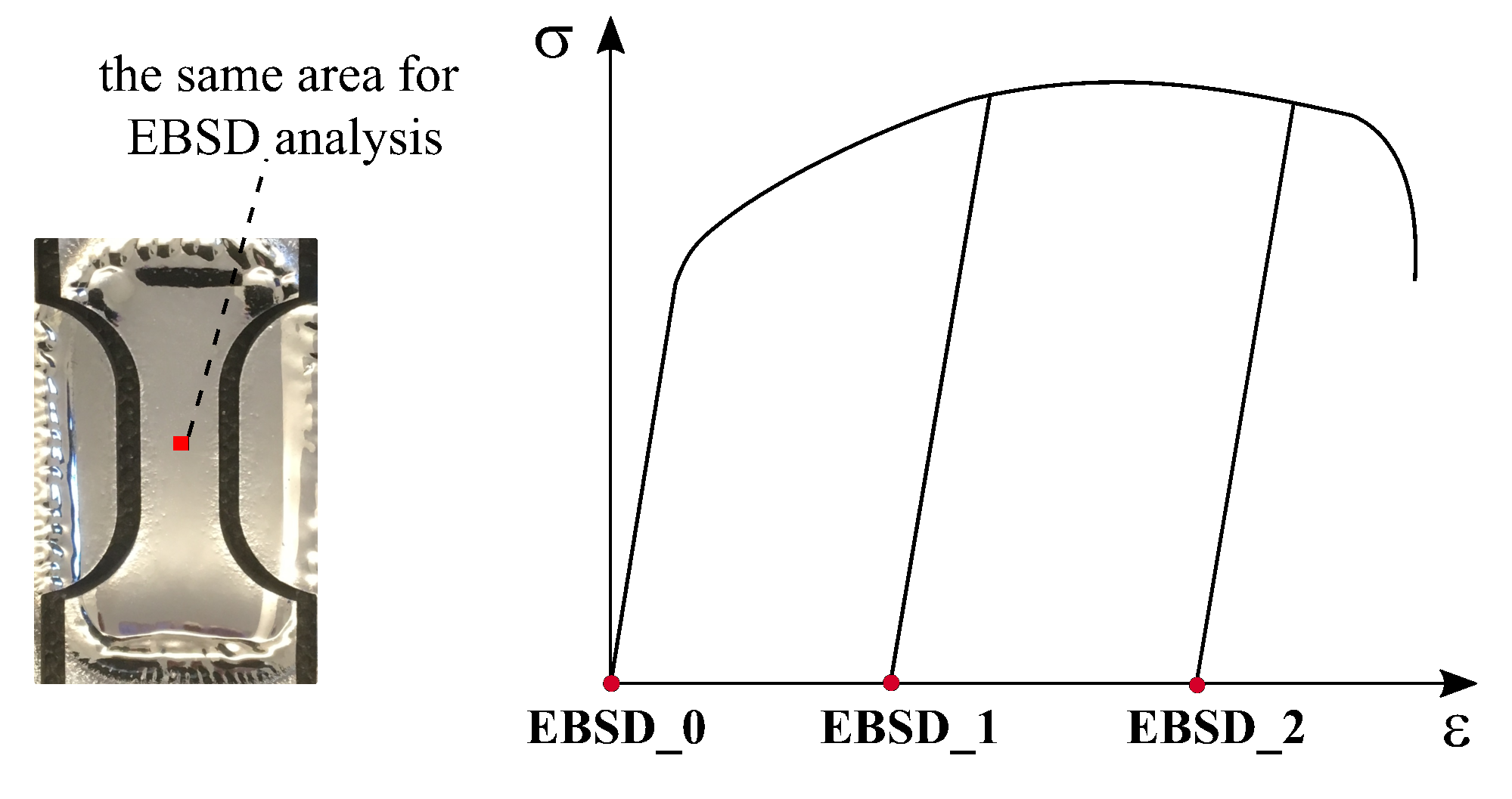}
		\caption{Tracking of crystallographic orientation in the same area during deformation: (left) analyzed region in the reference state; (right) schematic of sequential loading with intermediate unloadings and EBSD measurements.}
		\label{fig:EBSD_1st_method}
	\end{figure}

\begin{figure}[H]
		\centering
		\includegraphics[width=100mm]{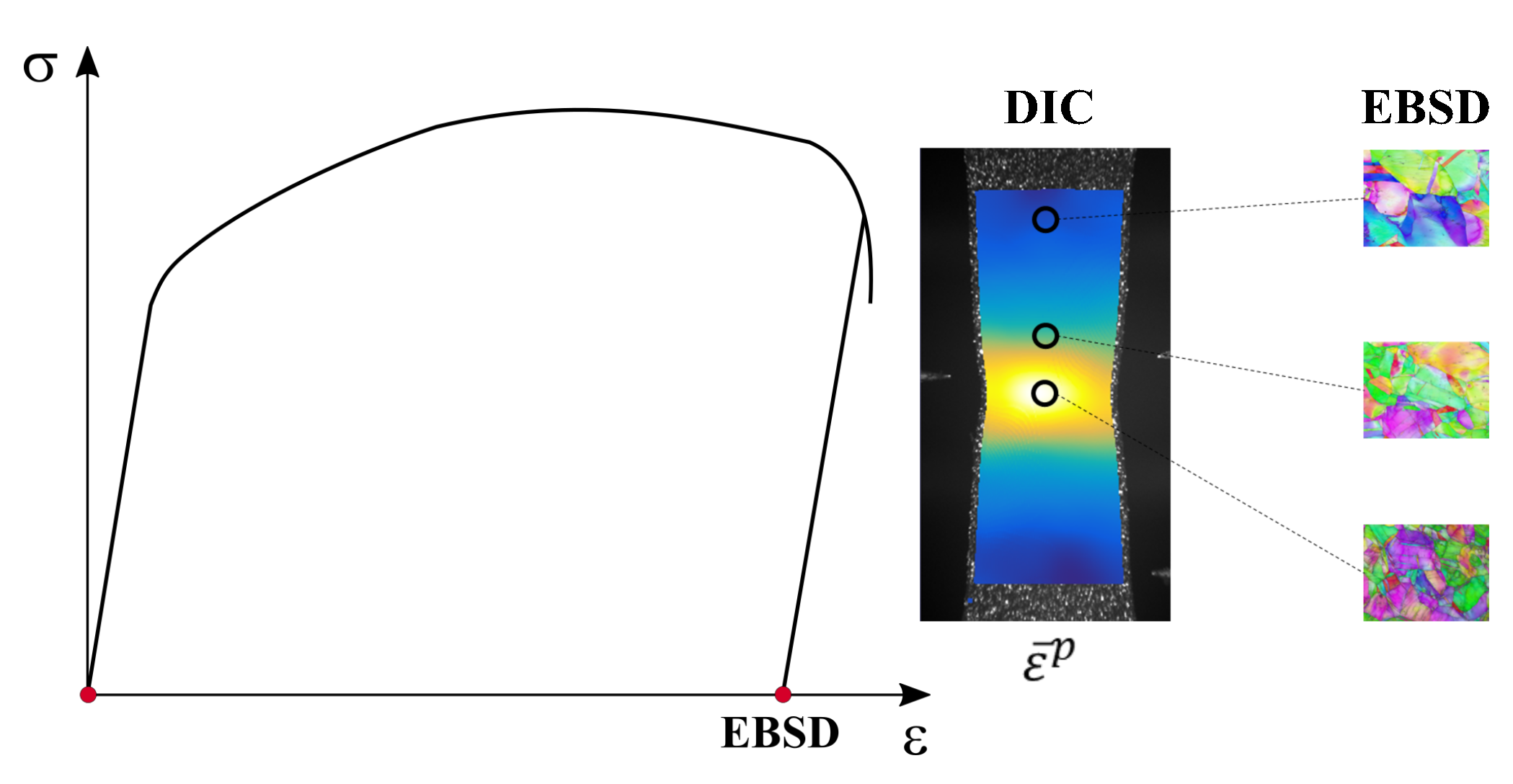}
		\caption{EBSD analysis of regions with different strain levels: (left) schematic of tensile test interrupted before fracture; (right) distribution of equivalent plastic strain with selected regions for EBSD measurements.}
		\label{fig:EBSD_2nd_method}
	\end{figure}

	Microstructural evolution during deformation was investigated using two complementary EBSD-based approaches.
	In the first approach, the evolution of crystallographic orientations within the same surface region was tracked at successive stages of deformation. The specimen surface was initially electropolished, and a rectangular region of interest was defined by marking its vertices using focused ion beam (FIB) milling. An EBSD map of this region was acquired in the undeformed state. The specimen was then subjected to uniaxial tensile loading with intermediate unloading steps, and EBSD measurements of the same region were performed after each loading-unloading cycle (Fig. \ref{fig:EBSD_1st_method}). This approach enabled direct tracking of orientation changes relative to the initial state. However, progressive surface roughening during deformation, combined with the near-surface nature of EBSD measurements, limited the accuracy of point-to-point spatial correspondence at higher strains.
	In the second approach, EBSD maps were acquired from different regions corresponding to known local strain levels. Specimens were deformed in uniaxial tension up to just prior to fracture (Fig. \ref{fig:EBSD_2nd_method}). The surface distribution of equivalent plastic strain ${\bar{\varepsilon}}^p$ was determined using DIC supported by a calibrated mechanical model \cite{musial2022field}. Based on this distribution, regions representing different strain levels were selected for EBSD analysis.
	To assess through-thickness homogeneity, specimens were sequentially electropolished to depths corresponding to one-quarter, one-half, and three-quarters of the sheet thickness, and EBSD measurements were performed on the resulting longitudinal cross-sections. This ensured that the analyzed regions represented bulk material and provided improved surface quality and spatial resolution. This approach enabled analysis of texture evolution as a function of strain, although direct comparison with the initial state was not possible.
	EBSD measurements were performed at an accelerating voltage of 25 kV in high-current mode using a 60 $\upmu$m aperture. Step sizes of 200, 100, and 50 nm were selected depending on the required spatial resolution. Data processing and visualization were carried out using ATEX software \cite{beausir2017analysis} and EDAX OIM Analysis.
	All EBSD orientation maps were primarily acquired on the RD-TD surface and subsequently transformed into the TD–ND reference frame to enable direct interpretation with respect to the tensile axis ($\parallel$RD). In addition, a complementary map was obtained directly in the RD-ND plane at an advanced stage of deformation within the necking region, providing insight into characteristic microstructural features from a through-thickness perspective.
	
	\section{Results and discussion}
	\label{results}
	\subsection{Macroscopically determined energy conversion}
	The results presented in this section are based on displacement and temperature fields measured during uniaxial tensile tests and analyzed using the thermomechanical framework proposed in~\cite{musial2022field}. Using this methodology, surface distributions of plastic work $w_p$ and energy dissipated $q_d$ were obtained, enabling evaluation of energy storage rate $Z$ at selected material points. The present study focuses on the dataset corresponding to a strain rate of $\dot{\varepsilon}=\SI{1e0}{\per\second}$. Based on the results presented in~\cite{musial2022field}, the evolution of the energy storage rate distribution, $Z$, was calculated using Thermocorr software \cite{ThermoCorr}. 
	\begin{figure}[H]
		\centering
		\includegraphics[width=160mm]{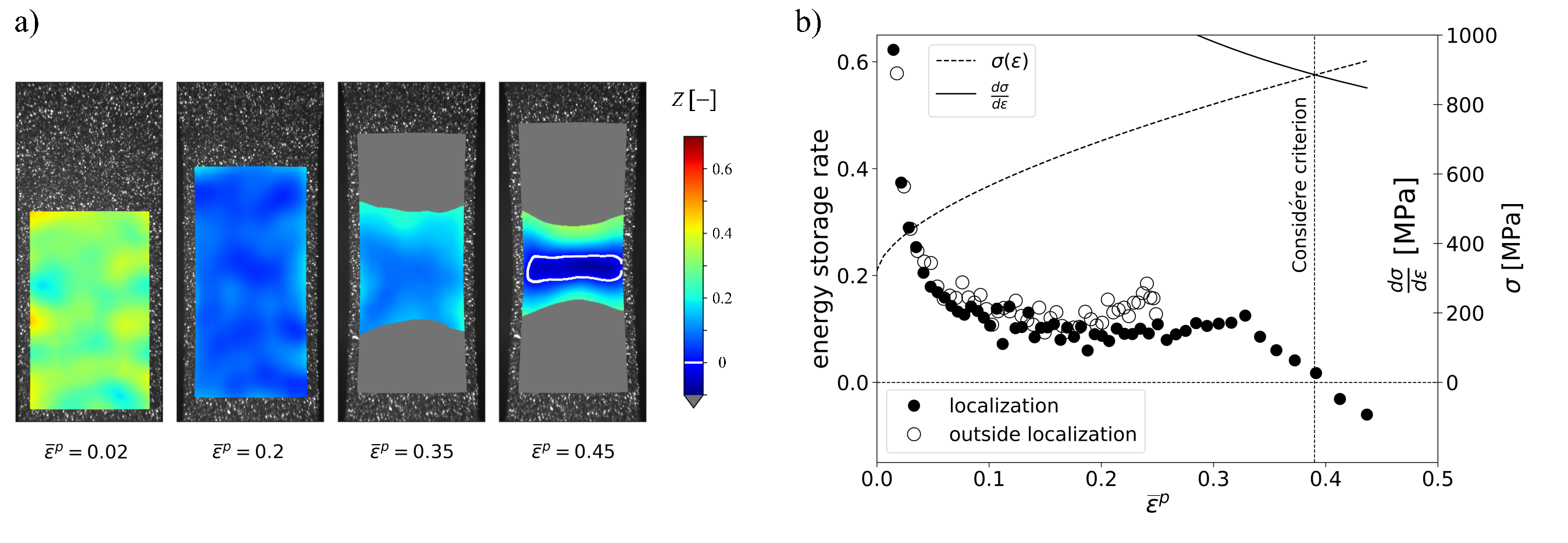}
		\caption{(a) Surface distributions of energy storage rate $Z$ at selected stages of deformation and (b) evolution of $Z$ at representative points located within and outside the localization zone \cite{musial2022field}, with the point corresponding to the Consid\'ere criterion indicated on the stress–strain curve.}
		\label{fig:Z_maps}
	\end{figure}
	Figure~\ref{fig:Z_maps}a shows the surface distributions of $Z$ at different stages of deformation, while Fig.~\ref{fig:Z_maps}b presents the evolution of $Z$ at two representative points: one located within the localization zone and the other outside it. In addition, Fig.~\ref{fig:Z_maps}b indicates the point on the stress–strain curve at which the Consid\'ere criterion is satisfied.
	
	At average plastic strains of $\varepsilon^p=0.02$ and $\varepsilon^p=0.2$, the $Z$ field remains spatially uniform, and the evolution of $Z$ at both points is similar, indicating macroscopically homogeneous deformation. At $\varepsilon^p=0.02$, relatively high values of $Z$ (on the order of 0.4) are observed. With increasing plastic strain, $Z$ decreases, reaching approximately 0.1–0.15 at $\varepsilon^p=0.2$. The distributions corresponding to $\varepsilon^p=0.35$ and $\varepsilon^p=0.44$ represent stages of strain localization. In these cases, $Z$ is evaluated only within the localized deformation zone, while regions outside - where plastic flow has ceased - are shown in gray. Accordingly, the evolution of $Z$ at the point outside the localization zone is plotted only up to the onset of localization. In the final stage, immediately prior to fracture, $Z$ decreases rapidly and reaches negative values, indicating that the material locally loses its ability to store energy before failure. The boundary of the region where the energy storage rate becomes non-positive ($Z \leq 0$) is marked by a white contour. A negative value of $Z$ implies that, for a given increment of plastic strain, the energy dissipated as heat exceeds plastic work, which can occur only if part of the energy stored during earlier stages of deformation is released. It is observed that the $Z = 0$ condition is reached at a plastic strain similar to that corresponding to the Consid\'ere criterion.
	\subsection{Microstructure evolution}
\begin{figure}[H]
		\centering
		\includegraphics[width=160mm]{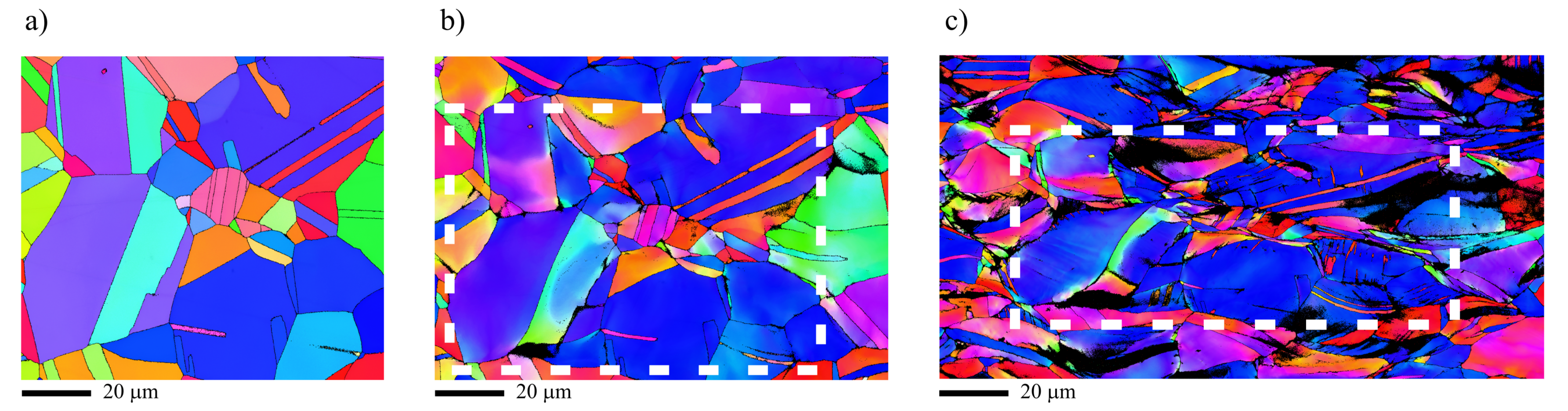}
		\caption{Orientation maps of the same area: (a) initial state, (b) macroscopically uniform deformation, (c) strain localization. The analyzed region is marked by a white dashed line in (b) and (c).}
		\label{fig:same_area_maps}
	\end{figure}

\begin{figure}[H]
		\centering
		\includegraphics[width=160mm]{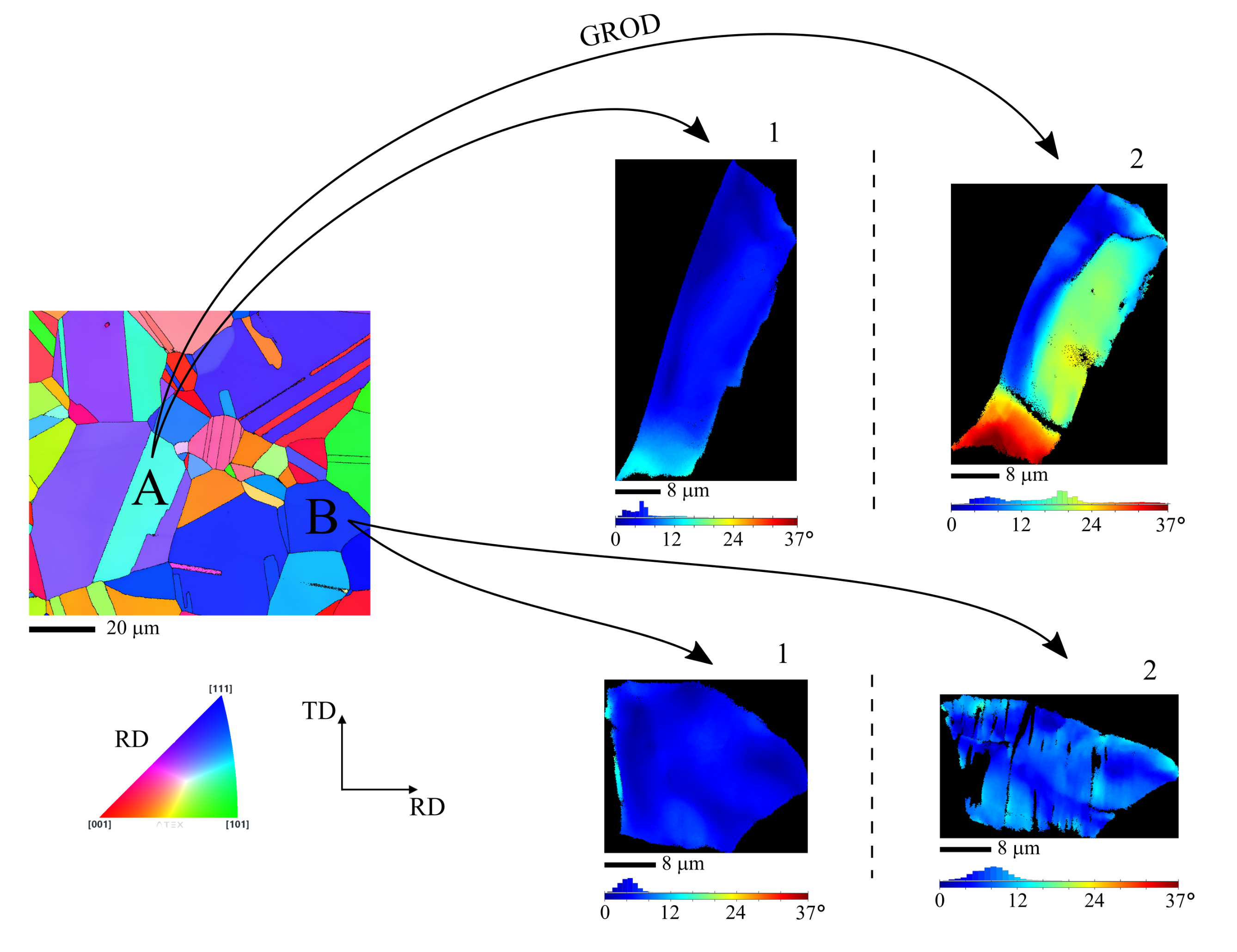}
		\caption{GROD distributions in grains A and B for (1) macroscopically uniform deformation and (2) strain localization.}
		\label{fig:GROD}
	\end{figure}

	Figure~\ref{fig:same_area_maps} presents inverse pole figure (IPF) maps of the same region at successive stages of deformation: the initial state (Fig.~\ref{fig:same_area_maps}a), macroscopically uniform deformation (Fig.~\ref{fig:same_area_maps}b), and the stage of plastic strain localization (Fig.~\ref{fig:same_area_maps}c). The maps were acquired with a step size of $\SI{100}{\nano\meter}$ and are color-coded with respect to the rolling direction (RD).
	Based on this sequence, the evolution of individual grain orientations was analyzed. Figure~\ref{fig:GROD} shows grain reference orientation deviation (GROD) distributions for two representative grains (A and B) with different initial orientations. GROD quantifies the local crystallographic misorientation within a grain relative to a selected reference orientation, here defined as the average grain orientation in the undeformed state. The GROD distributions are shown for two deformation stages: macroscopically uniform deformation (1) and strain localization (2).
	Grain A exhibits significantly higher GROD values than grain B, particularly at the stage of localized deformation. In grain A, GROD reaches up to \SI{37}{\degree}, reflecting pronounced intragranular lattice rotation and orientation heterogeneity. The resulting orientation gradients lead to the development of three distinct subregions, with increasingly well-defined boundaries as deformation localizes. In contrast, grain B shows a maximum GROD of \SI{15}{\degree}, with less pronounced internal heterogeneity.
	
\begin{figure}[H]
		\centering
		\includegraphics[width=160mm]{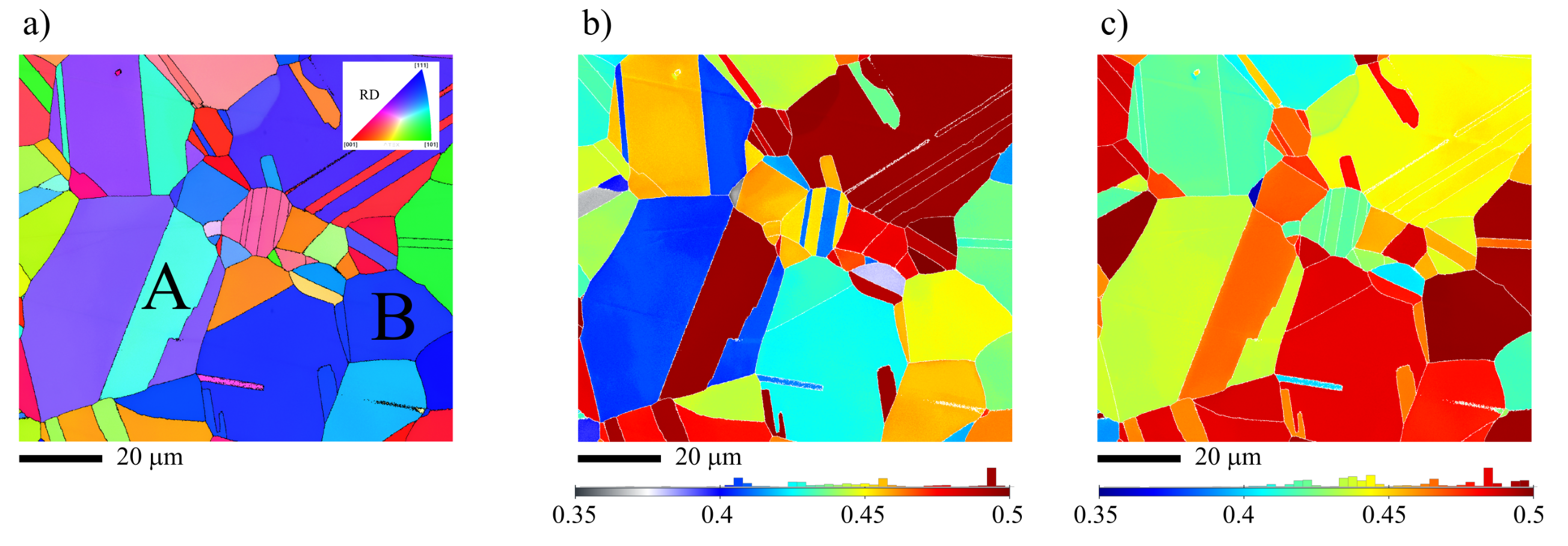}
		\caption{(a) Initial orientation map; (b) Schmid factor for slip ($SF^{slip}$); (c) Schmid factor for twinning ($SF^{twin}$).}
		\label{fig:SF_slip_twin}
	\end{figure}

\begin{figure}[H]
		\centering
		\includegraphics[width=160mm]{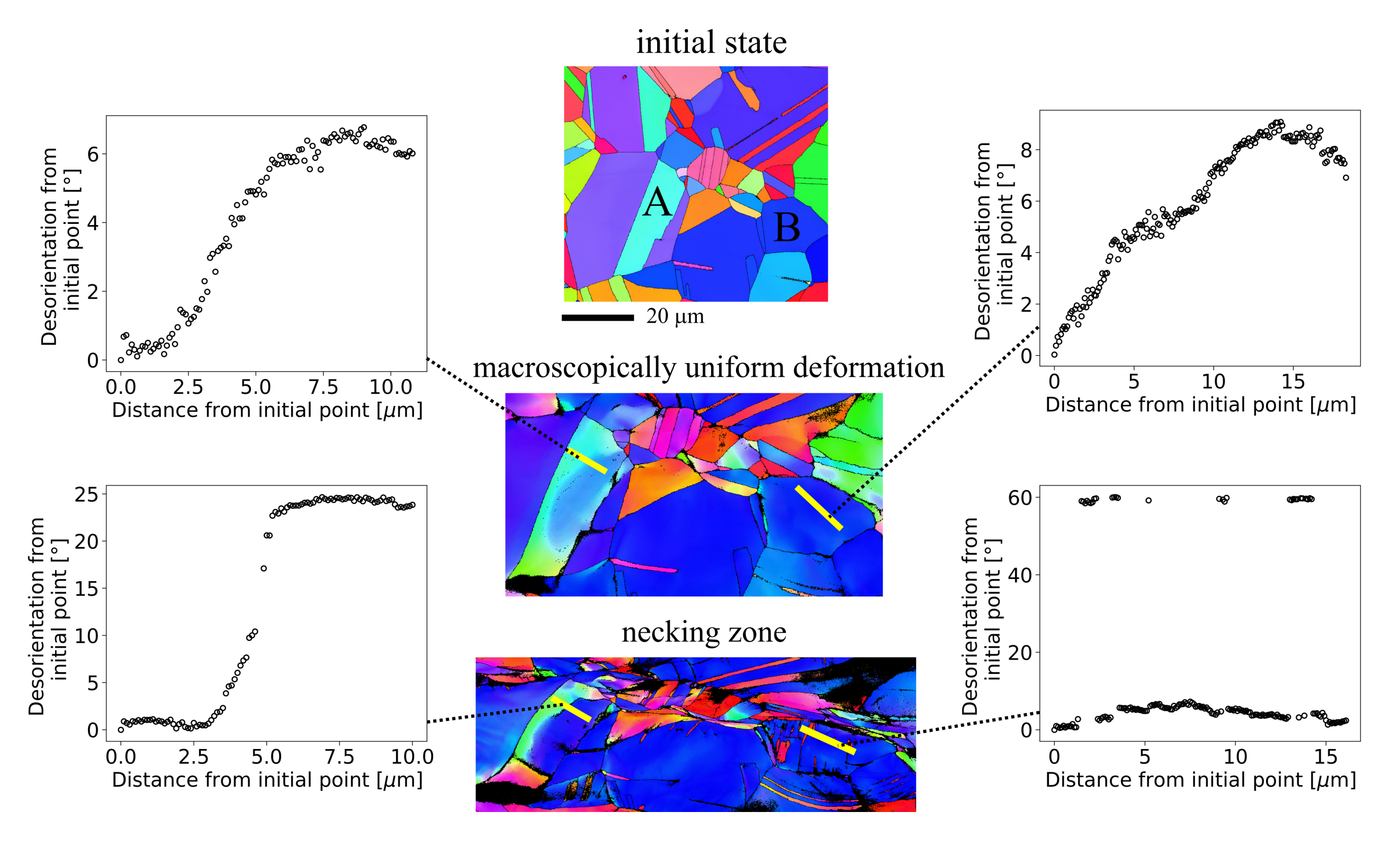}
		\caption{Disorientation profiles in grains A and B at different deformation stages.}
		\label{fig:profile_mis}
	\end{figure}

	These differences arise from the distinct initial crystallographic orientations of the grains, which influence the preferred deformation mechanisms. This tendency was evaluated using Schmid’s law, applied to both slip and twinning systems \cite{godet2006use}. Figure~\ref{fig:SF_slip_twin} shows the corresponding Schmid factor distributions for $\{111\}\langle110\rangle$ slip ($SF^{slip}$) and $\{111\}\langle112\rangle$ twinning ($SF^{twin}$) systems in the initial state.
	The Schmid factors were calculated considering 24 slip systems and 12 forward twinning systems. Grain A exhibits $SF^{slip}=0.495$ and $SF^{twin}=0.466$, indicating a preference for dislocation slip, whereas grain B ($SF^{slip}=0.45$, $SF^{twin}=0.497$) is more favorably oriented for twinning.
	Figure~\ref{fig:profile_mis} presents disorientation profiles for grains A and B at the stages of macroscopically uniform deformation and strain localization.
	In the macroscopically uniform regime, both grains deform predominantly by dislocation slip, with orientation changes of approximately \SI{7}{\degree} (grain A) and \SI{9}{\degree} (grain B).  During localized deformation, grain A continues to deform primarily by slip, but the orientation change increases to \SI{35}{\degree}, accompanied by the formation of sub-boundaries and fragmentation into distinct subgrains. The pronounced orientation gradients are consistent with increasing strain heterogeneity and the development of geometrically necessary dislocations (GNDs) to accommodate local lattice curvature. In contrast, grain B exhibits a transition toward deformation twinning following initial slip activity. Twin formation produces a characteristic lamellar twin–matrix structure and pronounced grain subdivision.
	
\begin{figure}[H]
		\centering
		\includegraphics[width=160mm]{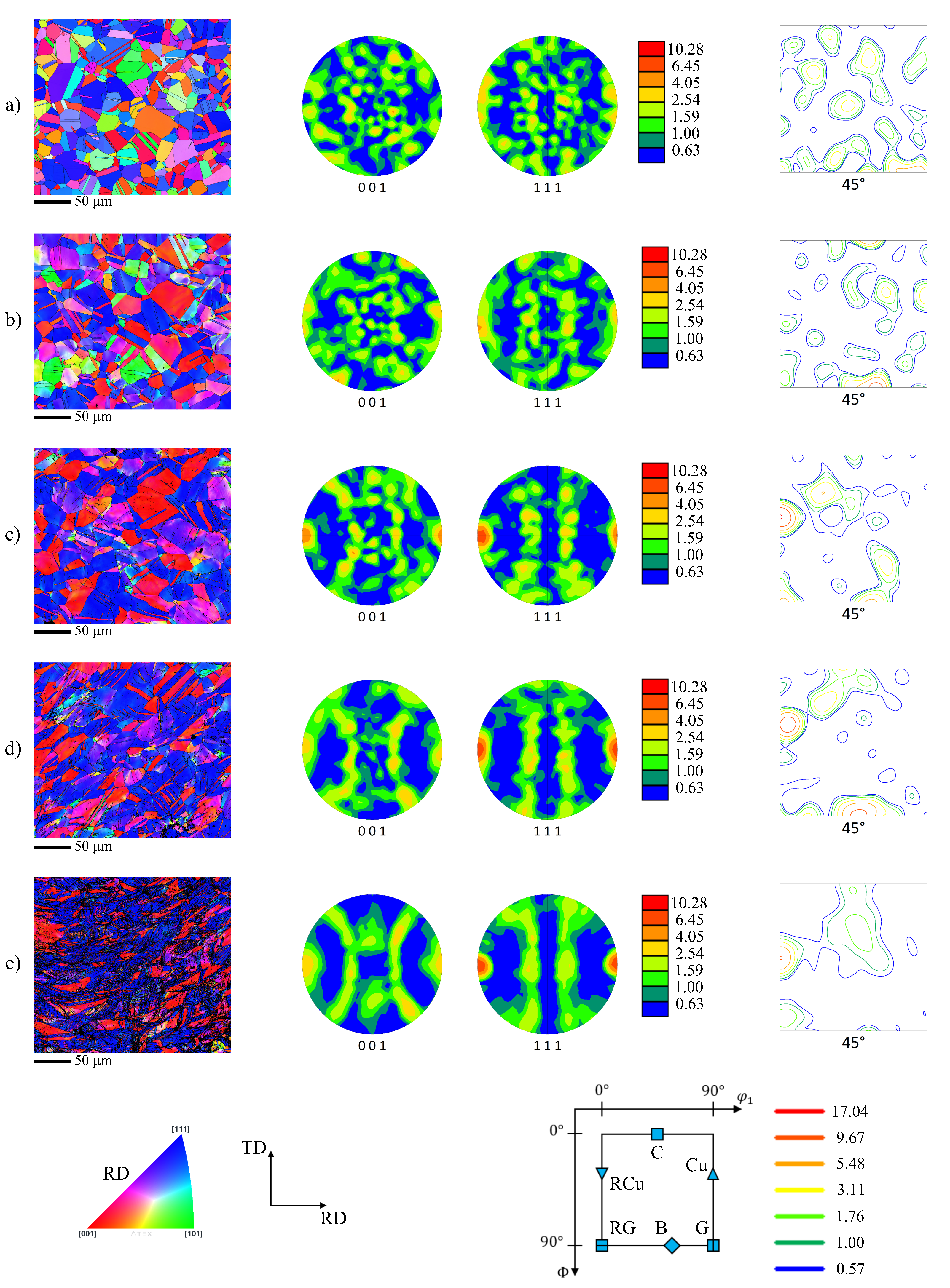}
		\caption{Orientation maps with respect to the tensile direction, the corresponding (111) and (100) pole figures and ODFs in the $\phi_{2}=\SI{45}{\degree}$ section for the successive values of the equivalent plastic strain ${\bar{\varepsilon}}^p$: (a) 0 (initial state), (b) 0.2, (c) 0.3, (d) 0.35, and (e) 0.45 (just before fracture).}
		\label{fig:texture}
	\end{figure}

	As described in Section~\ref{method}, the second experimental approach involved EBSD mapping of regions corresponding to known strain levels, enabling analysis of microtexture evolution. Measurements performed at different through-thickness locations showed consistent crystallographic orientation distributions, and mid-thickness maps were therefore selected as representative.
	Figure~\ref{fig:texture} presents orientation maps acquired with a step size of  $\SI{100}{\nano\meter}$, together with (111) and (100) pole figures for equivalent plastic strain levels of 0, 0.2, 0.3, 0.35, and 0.45 (just prior to fracture). The corresponding orientation distribution functions (ODFs) are shown in the $\phi_2=\SI{45}{\degree}$ section. The texture was calculated using OIM Analysis software based on a harmonic series expansion method with a series rank of 22. The texture is represented using Euler angles in the Bunge convention, and the ODF sections were computed with an angular resolution of $1^\circ$ in both $\phi_1$ and $\Phi$.
	In the initial state, the material exhibits a weak crystallographic texture inherited from rolling and subsequent heat treatment (Fig.~\ref{fig:texture}a). At ${\bar{\varepsilon}}^p=0.2$, the texture remains relatively weak (Fig.~\ref{fig:texture}b), although the Goss component (G, $\{110\}\langle100\rangle$), located near $\phi_1 = \SI{90}{\degree}, \Phi = \SI{90}{\degree}, \phi_2 = \SI{45}{\degree}$, diminishes and intensity develops near the Brass component (B, $\{110\}\langle112\rangle$), located at $\phi_1 = \SI{55}{\degree}, \Phi = \SI{90}{\degree}, \phi_2 = \SI{45}{\degree}$.
	
	For equivalent plastic strain between 0.2 and 0.3, deformation-induced components become more distinct and a redistribution of orientations occurs. The texture is dominated by Brass, Rotated Copper (RCu, $\{112\}\langle110\rangle$), and Rotated Goss (RG, $\{110\}\langle110\rangle$) components. Simultaneously, the Copper (C, $\{112\}\langle111\rangle$) and Goss (G) components progressively vanish, indicating their instability under continued tensile loading. Grain rotations align $\langle111\rangle$ and $\langle100\rangle$ directions with the tensile axis, leading to the development of a dual-fiber texture, with the $\langle111\rangle \parallel\mathrm{RD}$ fiber being dominant, consistent with the medium SFE of 310S steel \cite{dillamore1970stacking}. Similar dual-fiber evolution has been reported in another TWIP steels \cite{de2018twinning, barbier2009analysis}. At ${\bar{\varepsilon}}^p \approx 0.3$, the Brass component spreads toward the Rotated Copper orientation through a continuous lattice rotation of approximately $\SI{30}{\degree}$ about a $\langle111\rangle$ axis. This behavior is consistent with crystallographic slip on $\{111\}\langle110\rangle$ systems under relatively homogeneous deformation conditions.
	
	A qualitative transition occurs between ${\bar{\varepsilon}}^p=0.35$ and ${\bar{\varepsilon}}^p=0.45$. At ${\bar{\varepsilon}}^p=0.35$, the texture remains relatively compact and aligned with the $\langle111\rangle \parallel\mathrm{RD}$ fiber, indicating that deformation is still governed by planar slip and twinning-assisted rotation. In contrast, at ${\bar{\varepsilon}}^p=0.45$ (necking regime), the Brass component becomes diffuse, reflecting orientation fragmentation associated with strong local lattice curvature and grain subdivision. The increased intensity near the Rotated Goss component indicates a partial transition toward the $\langle100\rangle \parallel\mathrm{RD}$ fiber. Such redistribution cannot be explained by homogeneous slip alone and points to the activation of localized shear mechanisms. The observed transition B $\rightarrow$ (RCu + RG) is therefore consistent with shear-band-mediated lattice rotation, as reported for FCC materials undergoing large strain deformation \cite{paul2003shear}.
	
\begin{figure}[H]
		\centering
		\includegraphics[width=160mm]{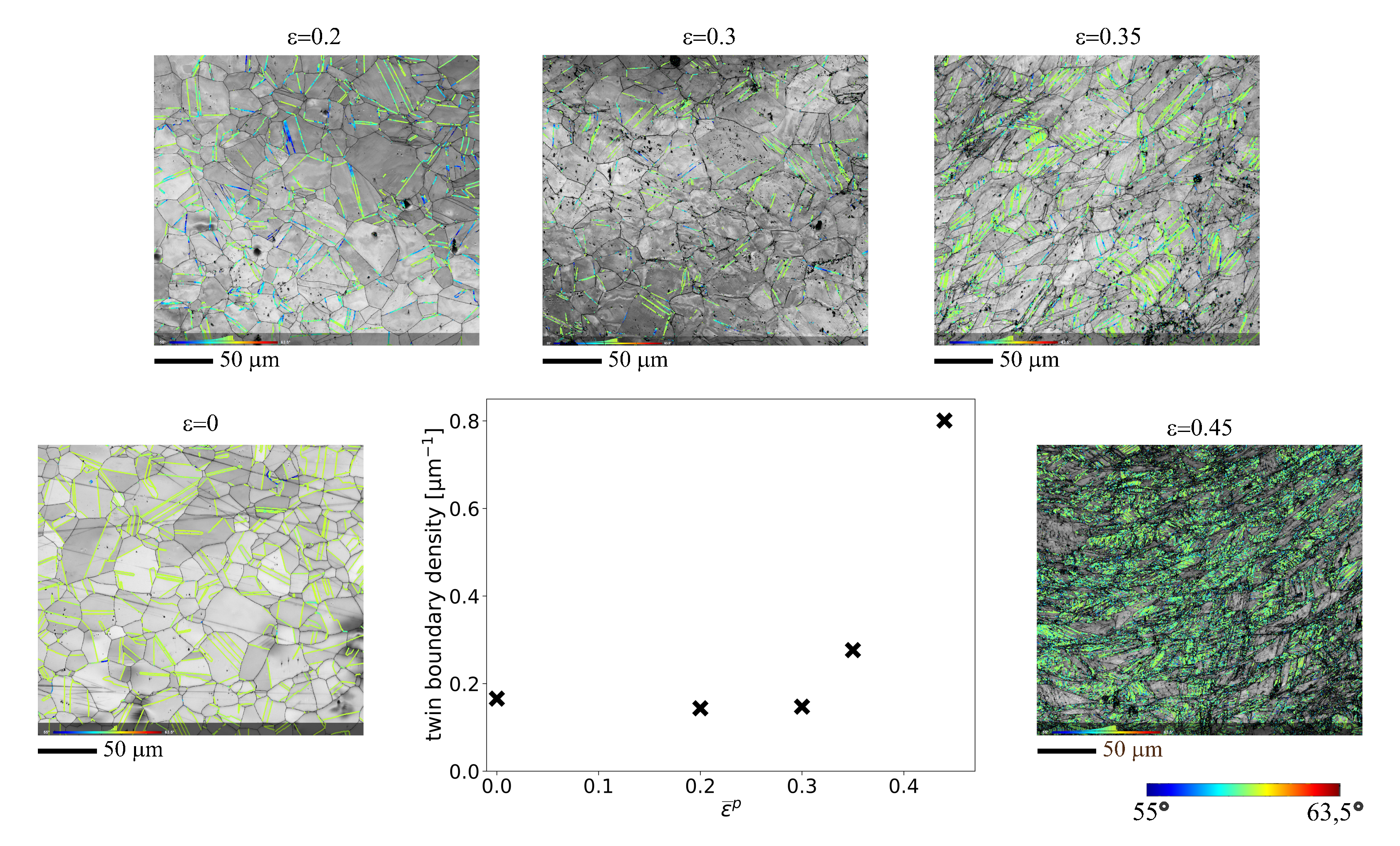}
		\caption{IQ maps for the analyzed strain levels, with marked CSL $\Sigma 3 - \SI{60}{\degree}\langle 111 \rangle$ twin boundaries, and a plot showing the relationship between twin boundary density and equivalent plastic strain.}
		\label{fig:twin_bound}
	\end{figure}

	This evolution is accompanied by a pronounced change in the deformation twin population. To quantify this change, the ($\Sigma3$) twin boundary density was evaluated as a function of plastic strain, as shown in Fig.~\ref{fig:twin_bound}. 
	In the reference state, only annealing twins are present. Up to ${\bar{\varepsilon}}^p=0.3$, the twin boundary density remains nearly constant (at the level of approximately $\SI{0.15}{\micro\meter}^{-1}$), indicating that deformation is primarily accommodated by dislocation slip. Beyond this strain, a rapid increase in twin boundary density is observed, indicating to the onset of deformation twinning within the resolution limits of EBSD. At ${\bar{\varepsilon}}^p=0.35$, numerous deformation twins form, predominantly in grains with $\langle111\rangle \parallel\mathrm{RD}$ orientation, resulting in a fine lamellar twin–matrix structure. At ${\bar{\varepsilon}}^p=0.45$, this lamellar morphology becomes pervasive with  twin boundary density exceeding $\SI{0.8}{\micro\meter}^{-1}$). Twinning does not introduce new texture components but redistributes the existing orientations: twins formed in $\langle111\rangle \parallel\mathrm{RD}$ grains promote the development of the $\langle100\rangle \parallel\mathrm{RD}$ fiber.
	
\begin{figure}[H]
		\centering
		\includegraphics[width=130mm]{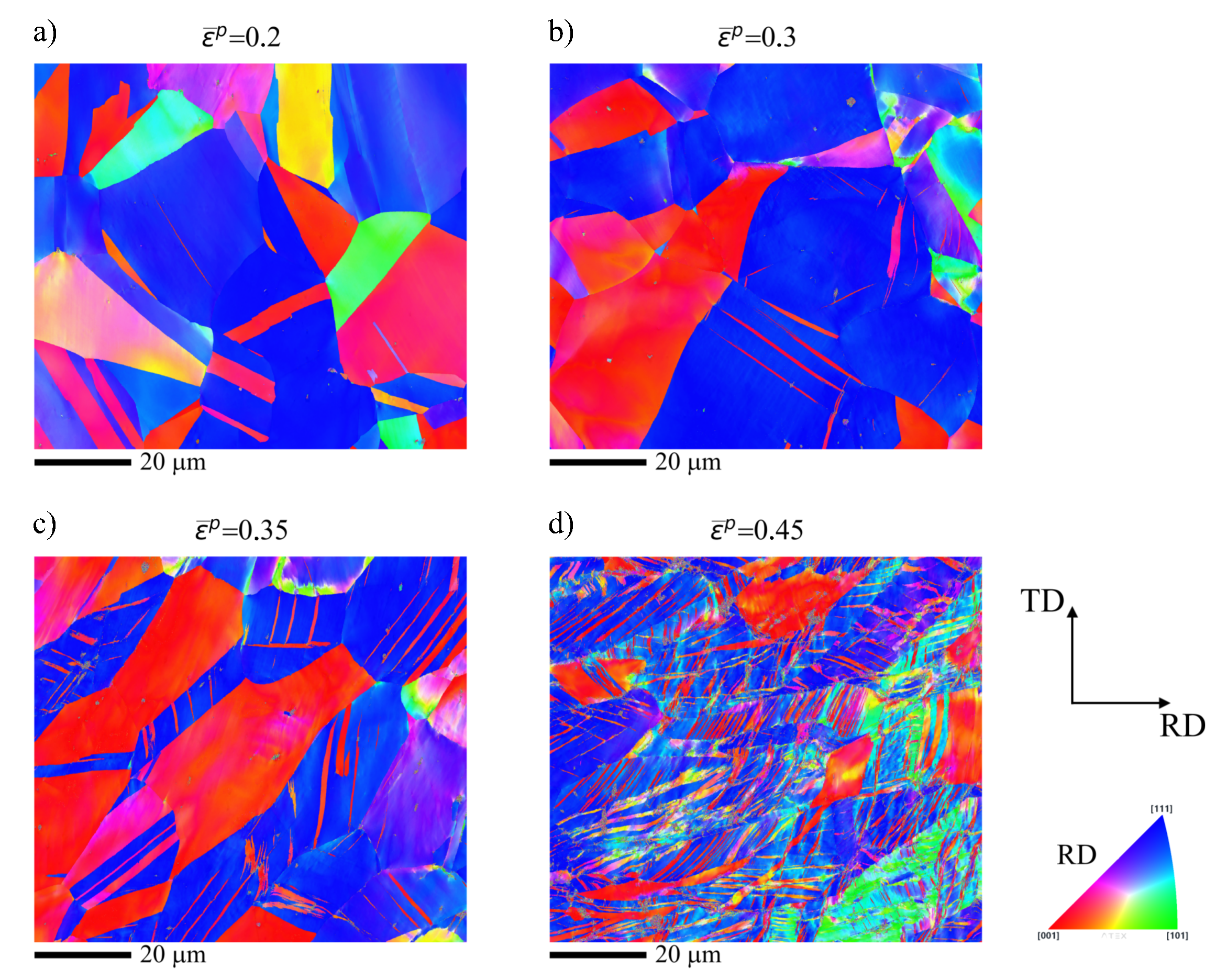}
		\caption{Orientation maps relative to the tensile direction, acquired at 3000x magnification with a $\SI{50}{\nano\meter}$ step size, for increasing equivalent plastic strain levels of (a) 0.2, (b) 0.3, (c) 0.35, and (d) 0.45 (just before specimen fracture).}
		\label{fig:mapy_twins_3kx}
	\end{figure}

	EBSD maps acquired with a finer step size of $\SI{50}{\nano\meter}$ (Fig.~\ref{fig:mapy_twins_3kx}) reveal a progressive increase in twin density and thickness with increasing strain. Moreover, deformation twins are not randomly distributed but form in crystallographically related variants that interact with one another. These variants are commonly organized into twin packets, in which multiple twin systems sharing a common $\{111\}$ habit plane operate within a single grain, leading to mutual impingement and progressive subdivision of the matrix into finer domains.
	The resulting lamellar structure markedly reduces the dislocation mean free path -- consistent with twinning-induced hardening (dynamic Hall–Petch effect) in TWIP steels \cite{YAO2021142044, frommeyer2003supra} --  and also enhances non-uniform strain distribution while promoting activation of multiple slip systems. These effects increase local strain incompatibility, thereby facilitating the development of localized deformation modes. 	
	
\begin{figure}[H]
		\centering
		\includegraphics[width=160mm]{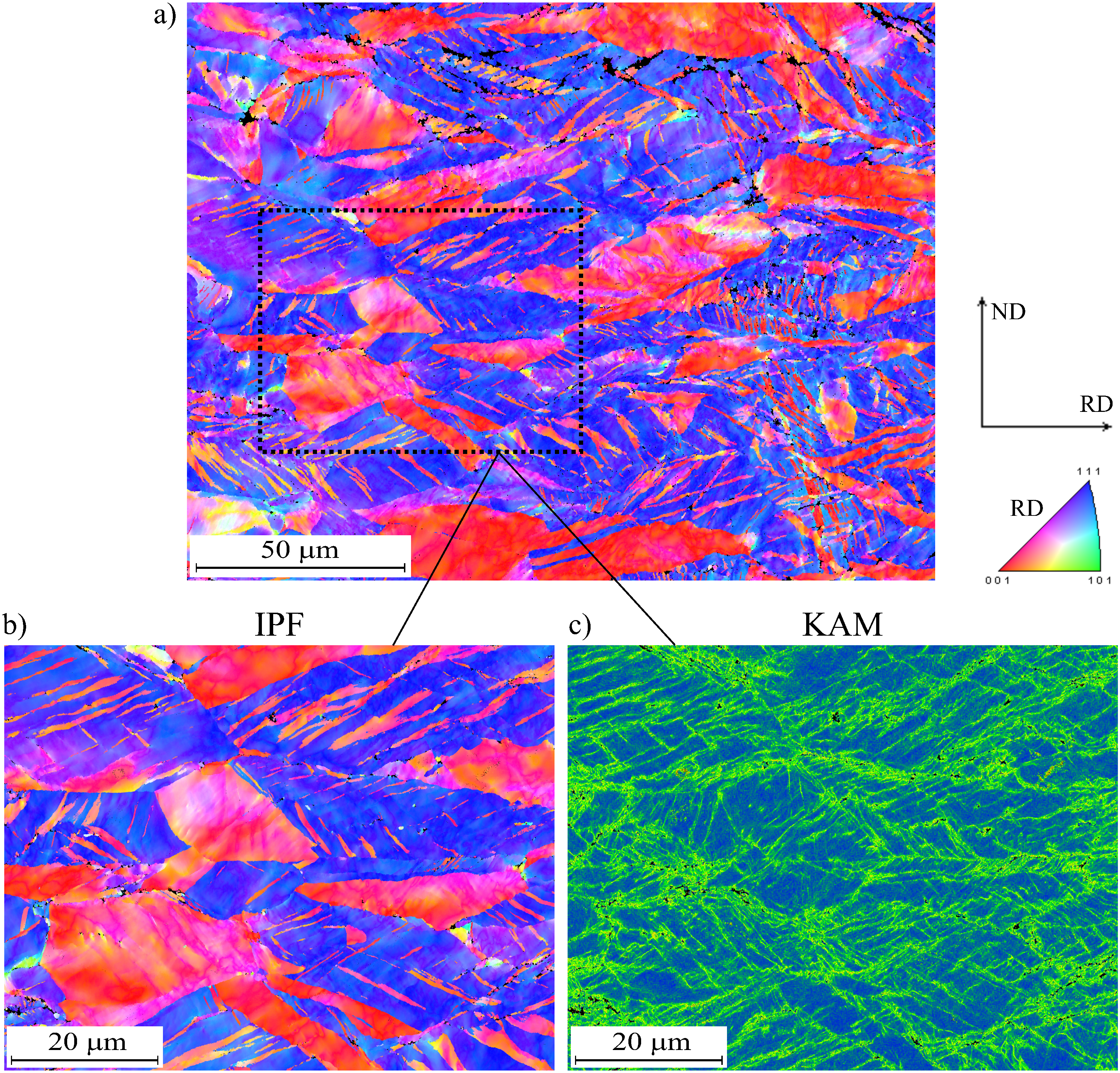}
		\caption{EBSD maps from the ND–RD section (longitudinal cross-section) in the necking zone obtained just prior to fracture: (a) overview IPF map (aligned with RD) acquired at 1500x with a $\SI{200}{\nano\meter}$ step size, (b) high-resolution IPF map (aligned with RD) of the subarea marked in (a), acquired at 3500x with a $\SI{50}{\nano\meter}$ step size, (c) corresponding Kernel Average Misorientation (KAM) map.}
		\label{fig:ND_RD_IPF_KAM}
	\end{figure}

	Figure \ref{fig:ND_RD_IPF_KAM} presents EBSD results obtained from the ND–RD section (longitudinal cross-section) in the necking zone, just prior to fracture. Overview and high-resolution maps are shown to highlight both the global morphology and local deformation features. Figure \ref{fig:ND_RD_IPF_KAM}a shows an IPF map (aligned with RD) acquired at lower magnification (1500x , step size 200 nm). A representative subarea, outlined by a black dashed contour, is examined in greater detail in Fig.~\ref{fig:ND_RD_IPF_KAM}b and c, which were acquired at higher magnification (3500x) with a refined step size of $\SI{50}{\nano\meter}$. Figure \ref{fig:ND_RD_IPF_KAM}b presents the corresponding IPF map (aligned with RD), while Fig.~\ref{fig:ND_RD_IPF_KAM}c shows the Kernel Average Misorientation (KAM) map. 
	The IPF maps reveal a pronounced longitudinal morphology characterized by a well-developed lamellar twin–matrix structure. Regions with $\langle111\rangle \parallel \mathrm{RD}$ orientations (associated with B and RCu components) coexist with $\langle100\rangle \parallel \mathrm{RD}$ regions (associated with the RG component and deformation twins), indicating significant orientation redistribution during deformation. The KAM map exhibits elevated local misorientation concentrated within twin-related domains, forming partially continuous high-strain channels aligned close to $\{111\}$ traces, consistent with incipient shear localization. These channel-like features appear quasi-parallel due to geometrical constraints imposed by the lamellar structure and are inclined at approximately $\SIrange{20}{45}{\degree}$ to the tensile axis, consistent with shear-band angles reported in the literature \cite{yeung1987shear, paul2003shear}. Within this framework, deformation twinning plays an indirect but critical role: rather than directly forming new texture components, it refines the microstructure and promotes strain heterogeneity, thus facilitating shear-band formation.
	
\begin{figure}[H]
		\centering
		\includegraphics[width=130mm]{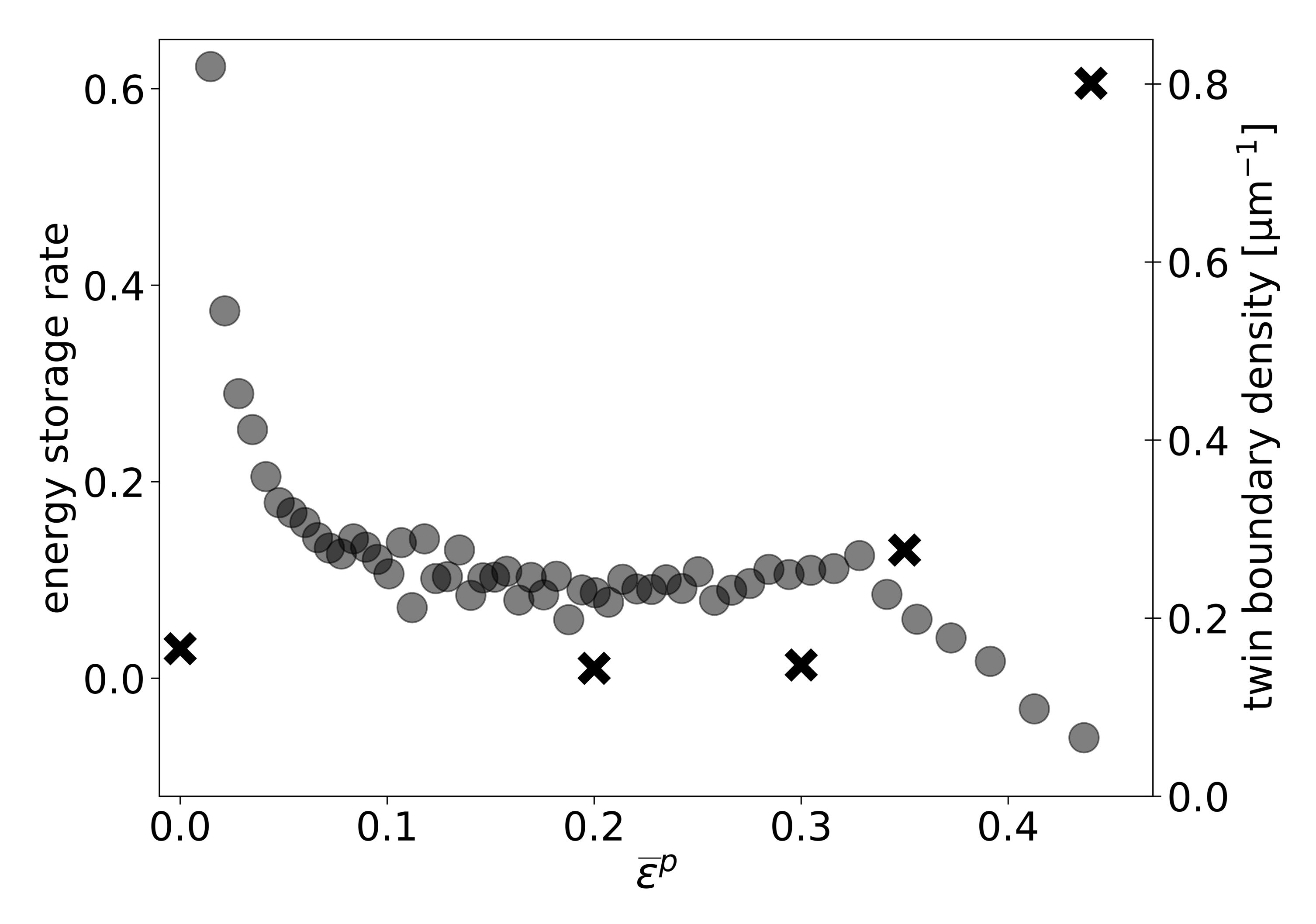}
		\caption{Variation in twin boundary density ($\times$) and energy storage rate ($\cdot$) as a function of equivalent plastic strain ${\bar{\varepsilon}}^p$}
		\label{fig:Z_vs_twin}
	\end{figure}

	The microstructural evolution described above has a direct impact on macroscopic plastic work partitioning. Figure~\ref{fig:Z_vs_twin} compares the evolution of the energy storage rate $Z$ with the twin boundary density. At low strains, $Z$ decreases rapidly while the twin boundary density remains nearly constant, indicating that energy storage is controlled by dislocation structure evolution. According to the LEDS theory \cite{hansen1986low}, this decrease reflects the formation of progressively lower-energy dislocation configurations. 
	In the strain range $0.15 \leq {\bar{\varepsilon}}^p \leq 0.3$, both $Z$ and twin boundary density remain relatively stable, consistent with predominantly homogeneous deformation controlled by dislocation slip. A distinct transition occurs beyond ${\bar{\varepsilon}}^p \approx 0.3$, where the rapid increase in twin boundary density coincides with a continued decrease in $Z$. 
	Recent results for HCP magnesium alloy \cite{maj2026plastic} have shown that twinning-dominated deformation can result in greater energy storage in the microstructure than slip-dominated deformation. 
	In the present TWIP steel, however, multiple concurrent mechanisms contribute to plastic work partitioning. Although deformation twinning introduces additional interfaces that can accommodate and store part of the plastic work, the development and refinement of the lamellar twin--matrix structure simultaneously promotes strain localization and enhances energy dissipation. Consequently, the additional energy storage associated with twin-boundary formation appears to be outweighed by the concurrent increase in energy dissipation, resulting in a continued decrease in $Z$ despite the pronounced increase in twin boundary density.
	
	In the necking regime, the microstructure is highly heterogeneous, comprising twin-free regions exhibiting strong orientation gradients and regions dominated by a fine lamellar twin-matrix structure (Fig.~\ref{fig:ND_RD_IPF_KAM}a,b), which, on the macroscopic scale, coincides with the area where $Z$ becomes non-positive (Fig.~\ref{fig:Z_maps}a). This behavior is consistent with earlier observations in austenitic steels, where strain localization and shear banding were associated with a loss of energy storage capacity \cite{oliferuk2007plastic}. Numerous studies have shown that layered microstructures composed of twins or dislocation arrangements promote shear band propagation \cite{duggan1978def, morii1985, leffers1990intra}. Within such bands, intense plastic deformation may activate recovery or dynamic recrystallization processes \cite{meyers1994evolution}, providing a mechanism for the local release of stored energy.
	
	\section{Summary and conclusions}
	
	A microstructural interpretation of plastic work partitioning during deformation of 310S TWIP steel was developed by correlating EBSD-based characterization of crystallographic orientation evolution with previously determined macroscopic energy conversion data. The analysis focused in particular on the stage of strain localization.
	
	The main conclusions are as follows:
	
	(i) Plastic deformation proceeds through a transition from dislocation-slip-dominated behavior at low strains to a regime of pronounced twinning activity beyond ${\bar{\varepsilon}}^p \approx 0.3$. This transition governs both grain-scale lattice rotation and the development of internal orientation gradients.
	
	(ii) Progressive deformation leads to the formation of a dual-fiber texture with dominant $\langle111\rangle \parallel\mathrm{RD}$ and secondary $\langle100\rangle \parallel\mathrm{RD}$ components. At advanced stages, texture evolution departs from homogeneous rotation paths and reflects increasing deformation heterogeneity.
	
	(iii) Deformation twinning does not introduce new texture components directly but refines the microstructure into a lamellar twin-matrix morphology. This refinement promotes non-uniform strain distribution, enhances local lattice curvature, and facilitates further crystallographic rotation under non-uniform deformation conditions.
	
	(iv) In the necking regime, the observed redistribution of texture components (B $\rightarrow$ RCu/RG) and the development of orientation fragmentation are consistent with shear-dominated deformation. The evolved lamellar structure and associated strain heterogeneities provide favorable conditions for shear bands initiation and propagation.
	
	(v) The evolution of microstructure is directly reflected in the macroscopic energy storage behavior. The initial decrease in the energy storage rate $Z$ is associated with dislocation rearrangement processes, while the onset of twinning does not lead to an increase in stored energy. Instead, during strain localization, the formation of a highly refined twin–matrix structure and intensified lattice rotations coincide with a reduction of $Z$ to zero or negative values.

	These findings demonstrate that plastic work partitioning in TWIP steels cannot be interpreted solely in terms of defect density. Instead, crystallographic reorientation, twinning-induced microstructural refinement, and strain localization collectively govern the transition toward deformation mechanisms that limit further energy storage and promote dissipation. The proposed approach provides a new microstructural perspective for interpreting plastic work partitioning in FCC materials where deformation twinning plays a dominant role.
	
	\section*{Data availability}
	
	The data supporting the findings of this study are available on Zenodo at
	\url{https://doi.org/10.5281/zenodo.18878199}.

	\section*{Acknowledgements}
	
	No funding was received for this research.\\The authors thank Leszek Urbański for performing the EBSD sample preparation, essential for the microstructural characterization presented in this work.
	
	\section*{Competing Interests}
	
 	The authors declare that they have no known competing financial interests or personal relationships that could have appeared to influence the work reported in this paper.
	
	\bibliographystyle{elsarticle-num}
	\bibliography{bibliography}

@article{Remy1977,
	title = {Twinning and strain-induced F.C.C. → H.C.P. transformation in the FeMnCrC system},
	journal = {Materials Science and Engineering},
	volume = {28},
	number = {1},
	pages = {99-107},
	year = {1977},
	issn = {0025-5416},
	doi = {10.1016/0025-5416(77)90093-3},
	author = {L. Remy and A. Pineau}
}

@article{Ullrich2016,
	title = {Interplay of microstructure defects in austenitic steel with medium stacking fault energy},
	journal = {Materials Science and Engineering: A},
	volume = {649},
	pages = {390-399},
	year = {2016},
	issn = {0921-5093},
	doi = {10.1016/j.msea.2015.10.021},
	author = {C. Ullrich and R. Eckner and L. Krüger and S. Martin and V. Klemm and D. Rafaja}
}

@article{Karaman2000,
	title = {Deformation of single crystal Hadfield steel by twinning and slip},
	journal = {Acta Materialia},
	volume = {48},
	number = {6},
	pages = {1345-1359},
	year = {2000},
	issn = {1359-6454},
	doi = {10.1016/S1359-6454(99)00383-3},
	author = {I. Karaman and H. Sehitoglu and K. Gall and Y. I. Chumlyakov and H. J. Maier}
}

@article{yang2006dependence,
	title={Dependence of deformation twinning on grain orientation in a high manganese steel},
	author={Yang, P. and Xie, Q. and Meng, L. and Ding, H. and Tang, Z.},
	journal={Scripta Materialia},
	volume={55},
	number={7},
	pages={629--631},
	year={2006},
	doi = {10.1016/j.scriptamat.2006.06.004},
	publisher={Elsevier}
}

@article{Huang2017,
	author = {Huang, T. T. and Dan, W. J. and Zhang, W. G.},
	title = {Study on the Strain Hardening Behaviors of TWIP/TRIP Steels},
	journal = {Metallurgical and Materials Transactions A},
	volume = {48},
	pages = {4553–-4564},
	year = {2017},
	publisher = {Springer},
	doi = {10.1007/s11661-017-4245-0}
}

@article{brofman1978effect,
	title={On the effect of carbon on the stacking fault energy of austenitic stainless steels},
	author={Brofman, P. J. and Ansell, G. S.},
	journal={Metallurgical Transactions A},
	volume={9},
	number={6},
	pages={879--880},
	year={1978},
	doi = {doi.org/10.1007/BF02649799}
}

@article{Chen2022,
	title = {Revealing the oxidation mechanism of 310S stainless steel in supercritical water via high-resolution characterization},
	journal = {Corrosion Science},
	volume = {200},
	pages = {110212},
	year = {2022},
	issn = {0010-938X},
	doi = {10.1016/j.corsci.2022.110212},
	author = {K. Chen and L. Zhang and Z. Shen and X. Zeng}
}

@article{Grajcar2018,
	author = {Grajcar, A. and Kozłowska, A. and Topolska, S. and Morawiec, M.},
	title = {Effect of Deformation Temperature on Microstructure Evolution and Mechanical Properties of Low-Carbon High-Mn Steel},
	journal = {Advances in Materials Science and Engineering},
	volume = {2018},
	number = {1},
	pages = {7369827},
	doi = {10.1155/2018/7369827},
	year = {2018}
}

@article{lee2014effect,
	title={Effect of the strain rate on the TRIP--TWIP transition in austenitic Fe-12 pct Mn-0.6 pct C TWIP steel},
	author={Lee, S. and Estrin, Y. and De Cooman, B. C.},
	journal={Metallurgical and materials transactions A},
	volume={45},
	number={2},
	pages={717--730},
	year={2014},
	publisher={Springer},
	doi = {10.1007/s11661-013-2028-9}
}

@article{GUTIERREZURRUTIA20103552,
	title = {The effect of grain size and grain orientation on deformation twinning in a Fe–22wt.$\%$ Mn–0.6wt.$\%$ C TWIP steel},
	journal = {Materials Science and Engineering: A},
	volume = {527},
	number = {15},
	pages = {3552-3560},
	year = {2010},
	issn = {0921-5093},
	doi = {10.1016/j.msea.2010.02.041},
	author = {I. Gutierrez-Urrutia and S. Zaefferer and D. Raabe}
}

@article{RAHMAN2015247,
	title = {The effect of grain size on the twin initiation stress in a TWIP steel},
	journal = {Acta Materialia},
	volume = {89},
	pages = {247-257},
	year = {2015},
	issn = {1359-6454},
	doi = {10.1016/j.actamat.2015.02.008},
	author = {K. M. Rahman and V. A. Vorontsov and D. Dye}
}

@article{Field2024,
	author = {Field, D. M. and Magagnosc, D. J. and Hornbuckle, B.C. and Lloyd, J .T. and Limmer, K. R.},
	title = {Manipulation of the Stacking Fault Energy of a Medium-Mn Steel Through Temperature and Hierarchical Compositional Variation},
	journal = {Metallurgical and Materials Transactions A},
	volume = {55},
	pages = {161–-172},
	year = {2024},
	publisher = {Springer},
	doi = {10.1007/s11661-023-07239-x}
}

@article{Wang2022,
	author = {X. Wang  and R. Rodriguez De Vecchis  and C. Li  and H. Zhang  and X. Hu  and S. Sridar  and Y. Wang  and W. Chen  and W. Xiong },
	title = {Design metastability in high-entropy alloys by tailoring unstable fault energies},
	journal = {Science Advances},
	volume = {8},
	number = {36},
	pages = {eabo7333},
	year = {2022},
	doi = {10.1126/sciadv.abo7333}
}

@article{HaileYan2014,
	author = {H. Yan and X. Zhao and N. Jia, Y. Zheng and T. He},
	title = {Influence of Shear Banding on the Formation of Brass-type Textures in Polycrystalline fcc Metals with Low Stacking Fault Energy},
	publisher = {J. Mater. Sci. Technol.},
	year = {2014},
	journal = {Journal of Materials Sciences and Technology},
	volume = {30},
	number = {4},
	eid = {408},
	pages = {408-416},
	doi = {10.1016/j.jmst.2013.11.010}
}

@article{ELDANAF20002665,
	title = {Deformation texture transition in brass: critical role of micro-scale shear bands},
	journal = {Acta Materialia},
	volume = {48},
	number = {10},
	pages = {2665-2673},
	year = {2000},
	issn = {1359-6454},
	doi = {10.1016/S1359-6454(00)00050-1},
	author = {E. El-Danaf and S. R. Kalidindi and R. D. Doherty and C. Necker}
}

@ARTICLE{Paul2007,
	author = {Paul, H. and Driver, J. H. and Maurice, C. and Pi{\k{a}}tkowski, A.},
	title = "{The role of shear banding on deformation texture in low stacking fault energy metals as characterized on model Ag crystals}",
	journal = {Acta Materialia},
	year = {2007},
	volume = {55},
	number = {2},
	pages = {575-588},
	doi = {10.1016/j.actamat.2006.08.051}
}

@article{titchener1958stored,
	title={The stored energy of cold work},
	author={Titchener, A. L. and Bever, M. B.},
	journal={Progress in Metal Physics},
	volume={7},
	pages={247--338},
	year={1958},
	publisher={Elsevier},
	doi = {10.1016/0079-6425(73)90001-7}
}

@article{taylor1934latent,
	title={The latent energy remaining in a metal after cold working},
	author={Taylor, G. I. and Quinney, H},
	journal={Proceedings of the Royal Society of London. Series A, Containing Papers of a Mathematical and Physical Character},
	volume={143},
	number={849},
	pages={307--326},
	year={1934},
	publisher={The Royal Society London},
	doi = {DOI:10.1098/rspa.1934.0004}
}

@article{wolfenden1971energy,
	title={The energy stored in polycrystalline copper deformed at room temperature},
	author={Wolfenden, A.},
	journal={Acta Metallurgica},
	volume={19},
	number={12},
	pages={1373--1377},
	year={1971},
	publisher={Elsevier},
	doi = {10.1016/0001-6160(71)90075-7}
}

@article{chrysochoos1989plastic,
	title={Plastic and dissipated work and stored energy},
	author={Chrysochoos, A. and Maisonneuve, O. and Martin, G. and Caumon, H. and Chezeaux, J. C.},
	journal={Nuclear Engineering and Design},
	volume={114},
	number={3},
	pages={323--333},
	year={1989},
	publisher={Elsevier},
	doi = {10.1016/0029-5493(89)90110-6}
}

@article{oliferuk1995effect,
	title={Effect of the grain size on the rate of energy storage during the tensile deformation of an austenitic steel},
	author={Oliferuk, W. and \'{S}wi\k{a}tnicki, W. A. and Grabski, M. W.},
	journal={Materials Science and Engineering: A},
	volume={197},
	number={1},
	pages={49--58},
	year={1995},
	publisher={Elsevier},
	doi = {10.1016/0921-5093(94)09766-6}
}

@article{oliferuk2009stress,
	title={Stress--strain curve and stored energy during uniaxial deformation of polycrystals},
	author={Oliferuk, W. and Maj, M.},
	journal={European Journal of Mechanics-A/Solids},
	volume={28},
	number={2},
	pages={266--272},
	year={2009},
	publisher={Elsevier},
	doi = {10.1016/j.euromechsol.2008.06.003}
}

@article{oliferuk1985energy,
	title={Energy storage during the tensile deformation of armco iron and austenitic steel},
	author={Oliferuk, W. and Gadaj, S. P. and Grabski, M. W.},
	journal={Materials Science and Engineering: A},
	volume={70},
	pages={131--141},
	year={1985},
	publisher={Elsevier},
	doi = {10.1016/0025-5416(85)90274-5}
}

@article{li2016local,
	title={Local experimental investigations of the thermomechanical behavior of a coarse-grained aluminum multicrystal using combined DIC and IRT methods},
	author={Li, L. and Muracciole, J. M. and Waltz, L. and Sabatier, L. and Barou, F. and Wattrisse, B.},
	journal={Optics and Lasers in Engineering},
	volume={81},
	pages={1--10},
	year={2016},
	publisher={Elsevier},
	doi = {10.1016/j.optlaseng.2016.01.001}
}

@article{louche2001thermal,
	title={Thermal and dissipative effects accompanying L{\"u}ders band propagation},
	author={Louche, H. and Chrysochoos, A.},
	journal={Materials Science and Engineering: A},
	volume={307},
	number={1-2},
	pages={15--22},
	year={2001},
	publisher={Elsevier},
	doi = {10.1016/S0921-5093(00)01975-4}
}

@article{wang2017kinematic,
	title={Kinematic and thermal characteristics of L{\"u}ders and Portevin-Le Ch{\^a}telier bands in a medium Mn transformation-induced plasticity steel},
	author={Wang, X. G. and Wang, L. and Huang, M. X.},
	journal={Acta Materialia},
	volume={124},
	pages={17--29},
	year={2017},
	publisher={Elsevier},
	doi = {10.1016/j.actamat.2016.10.069}
}

@article{bodelot2009experimental,
	title={Experimental setup for fully coupled kinematic and thermal measurements at the microstructure scale of an AISI 316L steel},
	author={Bodelot, L. and Sabatier, L. and Charkaluk, E. and Dufr{\'e}noy, P.},
	journal={Materials Science and Engineering: A},
	volume={501},
	number={1-2},
	pages={52--60},
	year={2009},
	publisher={Elsevier},
	doi ={10.1016/j.msea.2008.09.053}
}

@article{chrysochoos2009fields,
	title={Fields of stored energy associated with localized necking of steel},
	author={Chrysochoos, A. and Wattrisse, B. and Muracciole, J.-M. and El Ka{\"\i}m, Y.},
	journal={Journal of Mechanics of Materials and Structures},
	volume={4},
	number={2},
	pages={245--262},
	year={2009},
	publisher={Mathematical Sciences Publishers},
	doi = {10.2140/jomms.2009.4.241}
}

@article{knysh2015determination,
	title={Determination of the fraction of plastic work converted into heat in metals},
	author={Knysh, P. and Korkolis, Y. P.},
	journal={Mechanics of Materials},
	volume={86},
	pages={71--80},
	year={2015},
	publisher={Elsevier},
	doi = {10.1016/j.mechmat.2015.03.006}
}

@article{musial2022field,
	title={Field analysis of energy conversion during plastic deformation of 310S stainless steel},
	author={Musia{\l}, S. and Maj, M. and Urba{\'n}ski, L. and Nowak, M.},
	journal={International Journal of Solids and Structures},
	pages={111411},
	year={2022},
	publisher={Elsevier},
	doi= {10.1016/j.ijsolstr.2021.111411}
}

@article{oliferuk2007plastic,
	title={Plastic instability criterion based on energy conversion},
	author={Oliferuk, W. and Maj, M.},
	journal={Materials Science and Engineering: A},
	volume={462},
	number={1-2},
	pages={363--366},
	year={2007},
	publisher={Elsevier},
	doi = {10.1016/j.msea.2006.02.465}
}

@article{hansen1986low,
	title={Low energy dislocation structures due to unidirectional deformation at low temperatures},
	author={Hansen, N. and Kuhlmann-Wilsdorf, D.},
	journal={Materials Science and Engineering: A},
	volume={81},
	pages={141--161},
	year={1986},
	publisher={Elsevier},
	doi = {10.1016/0025-5416(86)90258-2}
}

@article{kuhlmann1987leds,
	title={LEDS: Properties and effects of low energy dislocation structures},
	author={Kuhlmann-Wilsdorf, D.},
	journal={Materials Science and Engineering: A},
	volume={86},
	pages={53--66},
	year={1987},
	publisher={Elsevier},
	doi = {10.1016/0025-5416(87)90442-3)}
}

@article{kuhlmann2001q,
	title={Q: Dislocations structures—how far from equilibrium? A: Very close indeed},
	author={Kuhlmann-Wilsdorf, D.},
	journal={Materials Science and Engineering: A},
	volume={315},
	number={1-2},
	pages={211--216},
	year={2001},
	publisher={Elsevier},
	doi = {10.1016/S0921-5093(01)01204-7}
}

@article{maj2026plastic,
  title={Plastic Work Partitioning During Slip-and Twinning-Dominated Deformation in AZ31B Magnesium Alloy},
  author={Maj, M. and Musia{\l}, S. and Nowak, M.},
  journal={Metallurgical and Materials Transactions A},
  pages={1--8},
  year={2026},
  publisher={Springer}
}

@article{beausir2017analysis,
	title={Analysis tools for electron and X-ray diffraction},
	author={Beausir, B. and Fundenberger, J. J.},
	journal={ATEX-software, www.atex-software.eu, Universit{\'e} de Lorraine-Metz},
	year={2017}
}

@article{godet2006use,
	title={Use of Schmid factors to select extension twin variants in extruded magnesium alloy tubes},
	author={Godet, S. and Jiang, L. and Luo, A. A. and Jonas, J. J.},
	journal={Scripta Materialia},
	volume={55},
	number={11},
	pages={1055--1058},
	year={2006},
	publisher={Elsevier},
	doi = {10.1016/j.scriptamat.2006.07.059}
}

@article{dillamore1970stacking,
	title={The stacking fault energy dependence of the mechanisms of deformation in fcc metals},
	author={Dillamore, I. L.},
	journal={Metallurgical Transactions},
	volume={1},
	pages={2463--2470},
	year={1970},
	publisher={Springer},
	doi = {10.1007/BF03038371}
}

@article{de2018twinning,
	title={Twinning-induced plasticity (TWIP) steels},
	author={De Cooman, B. C. and Estrin, Y. and Kim, S. K.},
	journal={Acta Materialia},
	volume={142},
	pages={283--362},
	year={2018},
	publisher={Elsevier},
	doi = {10.1016/j.actamat.2017.06.046}
}

@article{barbier2009analysis,
	title={Analysis of the tensile behavior of a TWIP steel based on the texture and microstructure evolutions},
	author={Barbier, D. and Gey, N. and Allain, S. and Bozzolo, N. and Humbert, M.},
	journal={Materials Science and Engineering: A},
	volume={500},
	number={1-2},
	pages={196--206},
	year={2009},
	publisher={Elsevier},
	doi = {10.1016/j.msea.2008.09.031}
}

@article{paul2003shear,
	title={Shear band microtexture formation in twinned face centred cubic single crystals},
	author={Paul, H. and Driver, J. H. and Maurice, C and Jasie{\'n}ski, Z.},
	journal={Materials Science and Engineering: A},
	volume={359},
	number={1-2},
	pages={178--191},
	year={2003},
	publisher={Elsevier},
	doi = {10.1016/S0921-5093(03)00335-6}
}

@article{YAO2021142044,
	title = {Static and dynamic Hall‒Petch relations in $\{332\} \langle113 \rangle$ TWIP Ti–15Mo alloy},
	journal = {Materials Science and Engineering: A},
	volume = {827},
	pages = {142044},
	year = {2021},
	issn = {0921-5093},
	doi = {10.1016/j.msea.2021.142044},
	author = {K. Yao and X. Min}
}

@article{frommeyer2003supra,
	title={Supra-ductile and high-strength manganese-TRIP/TWIP steels for high energy absorption purposes},
	author={Frommeyer, G. and Br{\"u}x, U. and Neumann, P.},
	journal={ISIJ international},
	volume={43},
	number={3},
	pages={438--446},
	year={2003},
	publisher={The Iron and Steel Institute of Japan},
	doi = {10.2355/isijinternational.43.438}
}

@article{duggan1978def,
	title={Deformation structures and textures in cold-rolled 70: 30 brass},
	author={Duggan, B. J. and Hatherly, M. and Hutchinson, W. B. and Wakefield, P. T.},
	journal={Metal Science},
	volume={12},
	number={8},
	pages={343--351},
	year={1978},
	publisher={Taylor \& Francis},
	doi = {10.1179/030634578790433909}
}

@article{morii1985,
	title={Development of shear bands in fcc single crystals},
	author={Morii, K. and Mecking, H. and Nakayama, Y.},
	journal={Acta Metallurgica},
	volume={33},
	number={3},
	pages={379--386},
	year={1985},
	publisher={Elsevier},
	doi = {10.1016/0001-6160(85)90080-X}
}

@article{leffers1990intra,
	title={Intra-and intergranular heterogeneities in the plastic deformation of brass during rolling},
	author={Leffers, T. and Bilde-S{\o}rensen, J. B.},
	journal={Acta Metallurgica et Materialia},
	volume={38},
	number={10},
	pages={1917--1926},
	year={1990},
	publisher={Elsevier},
	doi = {10.1016/0956-7151(90)90303-X}
}

@article{yeung1987shear,
	title={Shear band angles in rolled FCC materials},
	author={Yeung, W. Y. and Duggan, B. J.},
	journal={Acta Metallurgica},
	volume={35},
	number={2},
	pages={541--548},
	year={1987},
	publisher={Elsevier},
	doi = {10.1016/0001-6160(87)90259-8}
}

@article{meyers1994evolution,
	title={Evolution of microstructure and shear-band formation in $\alpha$-hcp titanium},
	author={Meyers, M. A. and Subhash, G. and Kad, B. K. and Prasad, L.},
	journal={Mechanics of Materials},
	volume={17},
	number={2-3},
	pages={175--193},
	year={1994},
	publisher={Elsevier},
	doi = {10.1016/0167-6636(94)90058-2}
}

@misc{ThermoCorr,
	title = "http://www.thermocorr.ippt.pan.pl",
}
	
\end{document}